# High pressure investigations on Hydrous Magnesium Silicate-Phase A using first principles calculations, H---H repulsion and O-H bond compression

H. K. Poswal, Surinder M Sharma and S. K. Sikka[†]

High pressure Physics Division, Bhabha Atomic Research Centre, Mumbai 400085, India
[†] Office of the Principal Scientific Adviser to the Government of India, Vigyan Bhawan Annexe, New Delhi 110011, India

We have carried out first principles structural relaxation calculations on the hydrous magnesium silicate Phase A ($Mg_7Si_2O_8(OH)_6$) under high pressures. Our results show that phase A does not undergo any phase transition upto ~ 45 GPa. We find that non-bonded H---H distance reaches a limiting value of 1.85 Å at about 45 GPa. The H---H repulsive strain releasing mechanism in Phase A is found to be dramatically different from the hydrogen bond bending one that was proposed by Hofmeister et al[1] for Phase B. It is based on the reduction of one of the O-H bond distances with compression.

**Introduction:** Dense hydrous magnesium silicates are believed to be major sources of water in subducting slabs and responsible for transport of water from surface into deep mantle of the Earth[2]. In most hydrous minerals, the hydrogen atom is bonded to oxygen as hydroxyl and which is usually hydrogen bonded. O-H---O hydrogen bonds exists in them in the whole range from weak to strong ones. A rough correlation seems to there between density (which can be translated into depth or pressure in the Earth) and the hydrogen bond strength[3]. At 0.1 MPa, the hydrous mineral phases,[4,5,6] phase A, phase B and superhydrous phase B contain two hydroxyl groups, hydrogen bonded to the same acceptor oxygen atom, as shown for phase A in Fig 1.

This arrangement is very unusual. The angle H1---O3---H2 is ~ 60° and contains short H1---H2 non bonded contacts. As determined by neutron and NMR spectroscopy[7] at 0.1 MPa, these distances (in Å units) are 2.10 Å, 1.86 Å, 1.93 Å for phase A, phase B and superhydrous phase B respectively. These are smaller than the value 2.4 Å, which is twice the van der Waals radius of the hydrogen atom and thus produce a repulsive strain in the structure. Application of pressure should further decrease these distances and intensify this repulsive strain. This may ultimately destabilise the structure of these compounds when H---H distance reaches a limiting value of about 1.85 Å, as is now known in many other compounds ( for a recent review of this effect see Sikka and Sharma[3] and references therein). However, before this limiting distance is reached upon compression, a hydrogen bond, which is a relatively weak interaction, may counter the effect of short H---H distances by compression of the O-H distance, bending of the hydrogen bond and disordering of the hydrogen atom sites. For phase B, Hofmeister et al[1] have postulated the hydrogen bond bending as the strain relieving mechanism. Normally, according to the correlation between O---H distance and angle H-O---O known at 0.1 MPa[8], this angle should decrease with increase of pressure. However, due to a wide range of H-O---H angle values possible at a given O---H length, the hydrogen bond may become more bent in some cases. This is what was deduced by Hofmeister et al[1] for phase B, using the infrared vibrational stretching frequency data and some assumptions about the O-H bond lengths, compressibility and maintaining the initial value of H---H distance of 1.84 Å (see Fig. 10 of ref. 1). For phase B, the calculated H-O---O angles for the two hydrogen bonds increase by 9 ± 1° and 11± 1° at 37 GPa.[1]

To throw more light on the H---H strain relieving mechanisms in these compounds, we have done first principles, density function theory (DFT) based, calculations on phase A as a first instance. This is the smallest structure ($Mg_7Si_2O_6(OH)_6$) among these three compounds. The structural parameters at 0.1 MPa were determined by Horiuchi et al[9] using single crystal x-ray diffraction and up to 9.4 GPa by Kudoh et al[10]. The hydrogen atom parameters were found by Kagi et al up to 3.2 GPa from a neutron powder diffraction study[4]. Phase A crystallizes in hexagonal space group $P6_3$ with 2 formula units in the unit cell [Fig. 1]. There are three Mg atoms, two Si atoms, six oxygen atoms and two hydrogen atoms in the asymmetric unit with 28 independent coordinates as variables. The pressure volume relation has been investigated by Crichton and Ross[11] up to 7.6 GPa and by Kudoh et al[10] up to 9.4 GPa. The compression behaviour of an iron bearing Phase A ($Mg_{6.85}Fe_{0.14}Si_{2.00}O_8(OH)_6$) has also been determined by x-ray powder diffraction up to 33GPa.[12] In this study, the c-axis appeared to show a sudden change in compression behaviour between 8 and 10 GPa. This may be due to a phase transition or due to the development of non-hydrostatic stress conditions in the diamond anvil cell owing to the freezing of pressure medium, silicone oil.[12]

Raman spectra of Phase A were measured up to 40 GPa by Liu et al.[13] They found phase A to undergo a reversible phase transition at 18 GPa. However, Hofmeister et al[1] did not find a transition up to 35 GPa and attributed the phase change detected by Liu et al to pressure induced hydration, because Liu et al employed water as the pressure medium in the diamond anvil cell. In contrast, Hofmeister et al did not use any pressure medium. Also, conflicting results were obtained for pressure variation (in cm$^{-1}$/GPa) of the two stretching frequencies. For example for 3518 cm$^{-1}$ mode, the $\partial v/\partial P$ determined by Liu[13] and Hofmister[1] were -0.058 and +0.31 cm$^{-1}$/GPa respectively, while for 3401cm$^{-1}$ mode this value is -0.016 cm$^{-1}$/GPa.[13] In many other hydrous minerals also positive pressure shifts have been observed for the stretch frequencies and sometimes have been attributed to H---H repulsions. Since, in many of them, e.g. chondroites, clinohumites etc.,[14] the hydrogen atoms are disordered, the effect can not be evaluated easily. Phase A, Phase B and superhydrous Phase B have ordered hydrogen atoms in them and thus are the ideal candidates for evaluation of the influence of H---H repulsion on the geometry of the hydrogen bonds.

**Computational Method:**

Structure relaxation calculations were performed employing the VASP (Vienna ab-initio simulation package) code[15-17] using density functional theory (DFT) within the framework of projected augmented wave method[18, 19]. We utilized the Perdiew Burke Enzorf generalized gradient approximation for exchange and correlation functional[20]. This functional has earlier been very successful in calculations of structural properties of hydrogen bonded systems.[21] The Brillouin zone was sampled by the Monkhorst-Pack scheme.[22]

Before undertaking the detailed calculations, we tested for the convergence with respect to the number of plane waves in the basis set and for **k** points sampling in the Brillouin zone (BZ). Using 15 **k** points in the irreducible Brillouin zone (IBZ) and $252\times10^{+3}$ plane waves, we obtained an energy convergence of $5\times10^{-3}$ eV. In the first cycle, lattice constants were kept fixed at the experimental values and all the ions were allowed to relax according to the calculated Hellman-Feynman forces, until the largest force component was less than $2\times10^{-3}$ eV/Å. This was followed by a series of fixed volume calculations in which ions as well as lattice parameters were allowed to relax. Calculations were started from the ambient structure determined by Horiuchi et al[9] and with hydrogen positions as found by Kagi et. al.[4] We have neglected the zero-point motion and thermal vibrations of the hydrogen atoms. The calculations were performed using 32 CPUs at the parallel station *Ajeya* at Bhabha Atomic Research Center.

**Results and Discussion:**

Table1 compares the calculated unit cell parameters at 0.1 MPa along with the experimental values. The agreement between the calculation and the experiments is reasonably good and is typical of density functional calculations.

Table 2 shows the comparison of experimentally measured atomic coordinates and the coordinates obtained after full relaxation calculation corresponding the experimental volume at the ambient conditions. Again, the match is reasonably good except for the H2 atom. This is not unexpected as the value of O4-H2 distance is only 0.89Å in neutron diffraction study of Kagi et al[4] The agreement at the volume of the experimental pressure of 9.4 GPa for the non hydrogen atoms is also shown in table 2 and is also good.

Figure 2 shows the experimental and theoretically predicted variation of $V/V_0$ with pressure. For the pure phase A the agreement is excellent upto the pressure for which the x-ray data is available ( ~9 GPa). The calculated results when fitted to third order Birch-Murnghan equation of state (EOS) give the zero pressure bulk modulus, $B_0$ = 96.2 GPa and its pressure derivative, $B_0'$ = 4.73, in reasonably good agreement with the experimentally measured values (see Table1)

DFT results do not indicate any subtle transition like the change in compressibility or axial ratio etc. This is more clearly reflected through the linearity of a plot of pressure–volume data in the universal EOS form given in Fig 3. The smooth behaviour of structural parameters as depicted in following figures is also in agreement with absence of any phase transition upto ~ 45 GPa.

Plot of H---H distance versus pressure is shown in Fig. 4 and it reaches the limiting distance of 1.85Å value at about 45 GPa. As mentioned above, to explain the pressure induced changes of hydroxyl stretching modes in phase B, Hofmeister et al argued that compression substantially increases the bond bending i.e. an increase in H-O---O angles. In phase A, at ambient conditions the angles H1-O2---O3 ($\delta1$) and H2-O4---

O3 (δ2) are ~ 6.8° and 1.2° respectively and these increase marginally by 3.3° and 2.6° respectively at 45 GPa. These changes are much smaller than the values given by Hofmeister for phase B. Therefore, the hydrogen bond bending does not appears to be the main mechanism for countering the H---H repulsion in phase A.

Further, we note that the angle H1---O3---H2 increases by almost 8° upto 45 GPa. The calculated variation is shown in Fig. 5. For phase B, Hofmeister et al had implied a reduction in this angle. In addition, the plot of O-H distances with pressure, given in Fig. 6, is very revealing. One of the distances (O4-H2) increases as expected from the correlation between O-H versus H---O distances.[23 and references therein] The other (O2-H1) decreases with pressure even though its H---O distance also decreases with pressure. This is unique, to the best of our knowledge, in that it is for first time a reduction has been seen in an O-H distance with compression. But for this, the H---H distance would have deceased faster and would have led to a higher repulsive strain. The decrease in O-H distance is only 0.0044 Å over the pressure range of 45 GPa. This is not amenable to measurements at present by neutron diffraction. However, there may a clue of this behaviour in spectroscopic measurements. In infrared and Raman techniques, this bond has been associated with the stretching frequency of 3400 cm$^{-1}$. Its pressure variation should be influenced by this decrease in O-H distance. In this context we note that Liu et al[13] had observed a very small variation of the frequency of this mode with pressure. However, more careful measurements are required to corroborate this.

It has been pointed out by Sikka and Sharma[3] that even at 0.1 MPa if the density of a hydrogen bonded substance is high, the acceptor coordination of the hydrogen atoms may be more than 1. The acceptor coordination is defined to be the number of acceptor atoms within the sum of the vdW radii of H and acceptor atom. Most of the hydrous phases seem to be having this property of having bifurcated or multi-centered hydrogen bonds. This was already recognised for phase A by Kagi et al[4] at 0.1 MPa. Fig. 7 illustrates this, where all oxygen atoms surrounding hydrogen atoms within 2.6Å are shown. Even if some of the contacts may not be called hydrogen bonds at lower pressures but under pressure they may become eligible as some of these distances will decrease, giving rise to multi-centered hydrogen bonds (compare plots (a) and (b) at ambient volume and at 45 GPa respectively for Phase A ). The hydrogen bonding for both the H1 and H2 atoms under compression is now of this type. This multi-centered nature of hydrogen bonds in Phase A may also contribute in relieving the H---H repulsion at high pressures.

To conclude, our first principles calculations show that the H---H strain relieving mechanism in phase A ($Mg_7Si_2O_8(OH)_6$) is quite distinct from the proposed mechanism in Phase B. In Phase A, the non-bonded H---H distance reach the limiting value by ~ 45 GPa and is accompanied by opening of H---O---H angle as well as compression of one of the O-H bond lengths.

# Figure Captions

Fig. 1: Structure of hydrous magnesium silicate - phase A projected along the c-axis. O3 is the common acceptor oxygen atom for two hydrogen bonds (O2-H1---O3, O4-H2---O3).

Fig. 2: $V/V_0$ as a function of pressure. The solid line represents the results of present calculations while open circles are from Crichton et al[11], triangles represent Fe bearing Phase A[12] and asterisk for Kudoh et al[10].

Fig. 3: Universal equation of state $\ln H = \ln[PX^2/\{3(1-X)\}] = \ln B_0 + (3/2)(B_0' - 1)(1-X)$, where $X = (V/V_0)^{1/3}$ and $B_0$ and $B_0'$ are bulk modulus and its pressure derivative.

Fig. 4: Computed non-bonded H---H distance as a function of pressures for phase A

Fig. 5: Computed variation of H1---O3---H2 angle as a function of pressure for phase A

Fig. 6: Calculated changes in O-H bond lengths as function of pressure. Solid circles represent O4-H2 and open circles denote O2-H1

Fig. 7: Oxygen co-ordination of the hydrogen atoms in the hydrogen bonds of phase A (a) at 2.2 GPa and (b) at 45 GPa. Inter-atomic distances are given in Å.

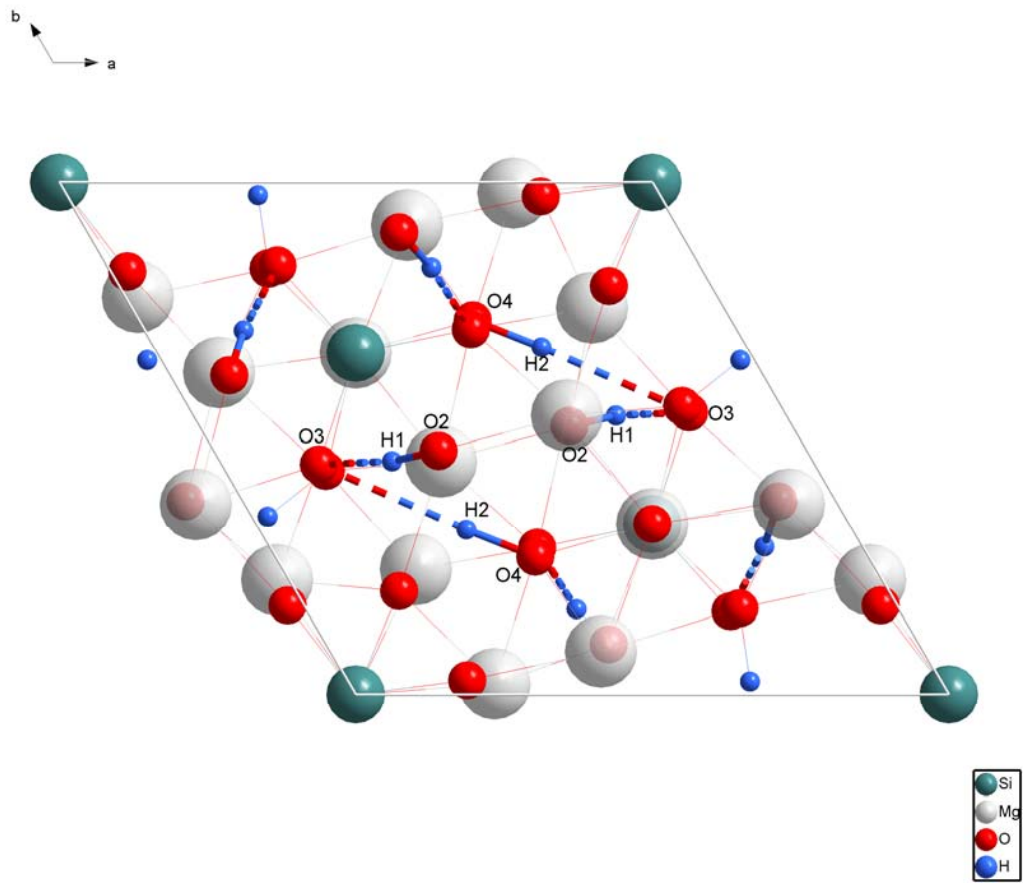

Figure 1

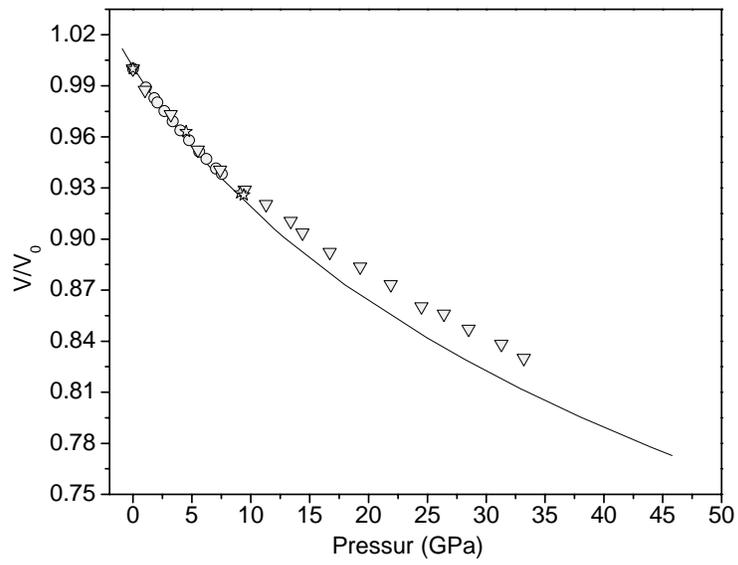

**Figure 2**

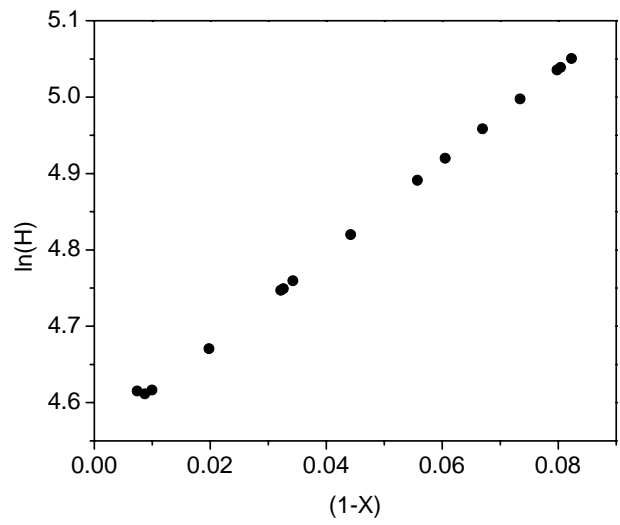

**Figure 3**

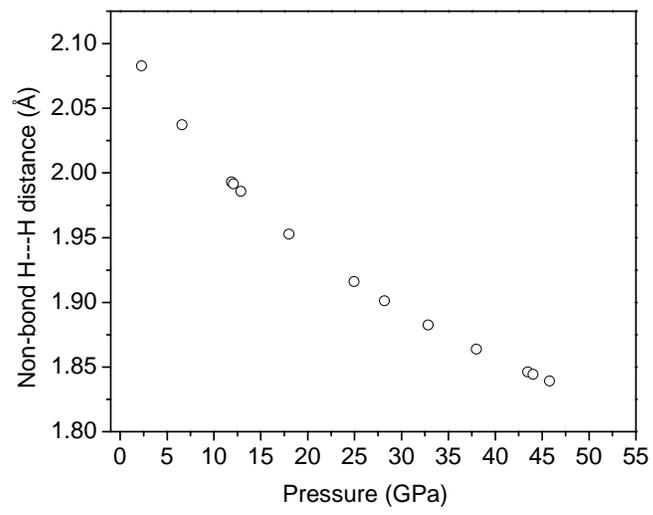

**Figure 4**

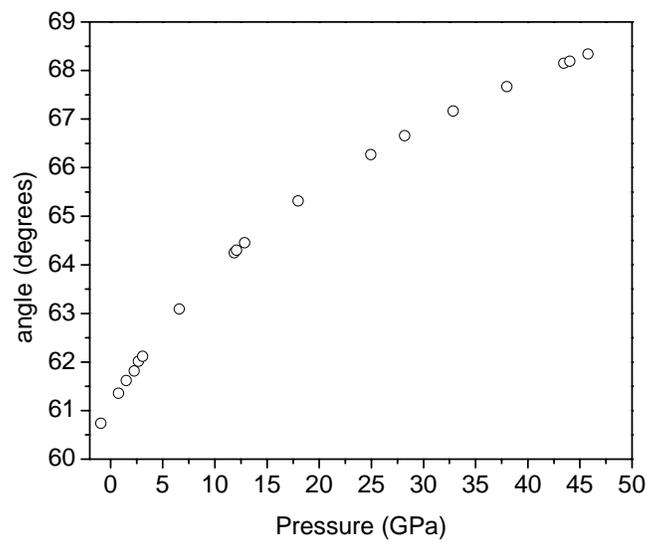

**Figure 5**

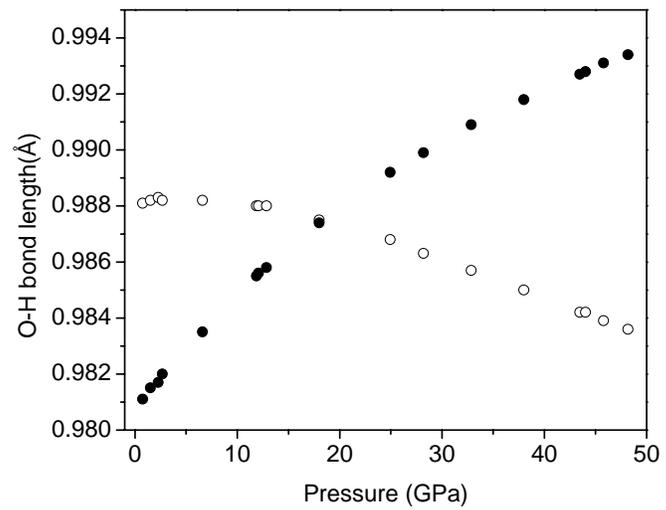

**Figure 6**

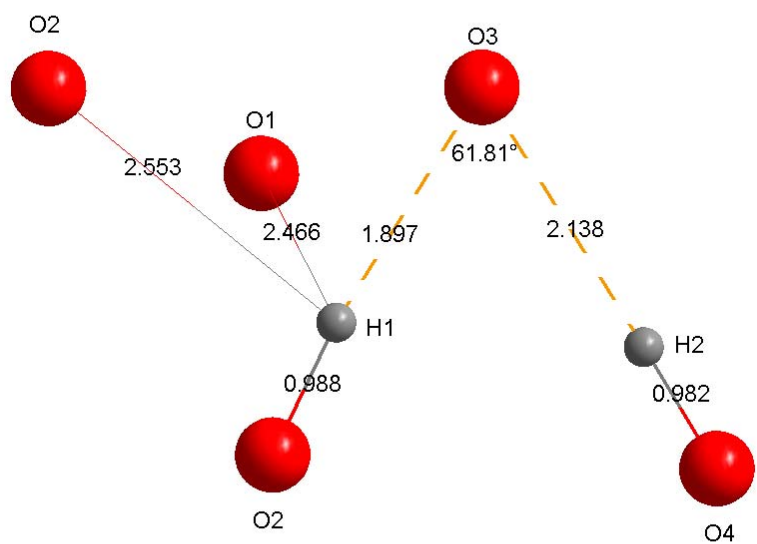

**(a)**

**Fig. 7**

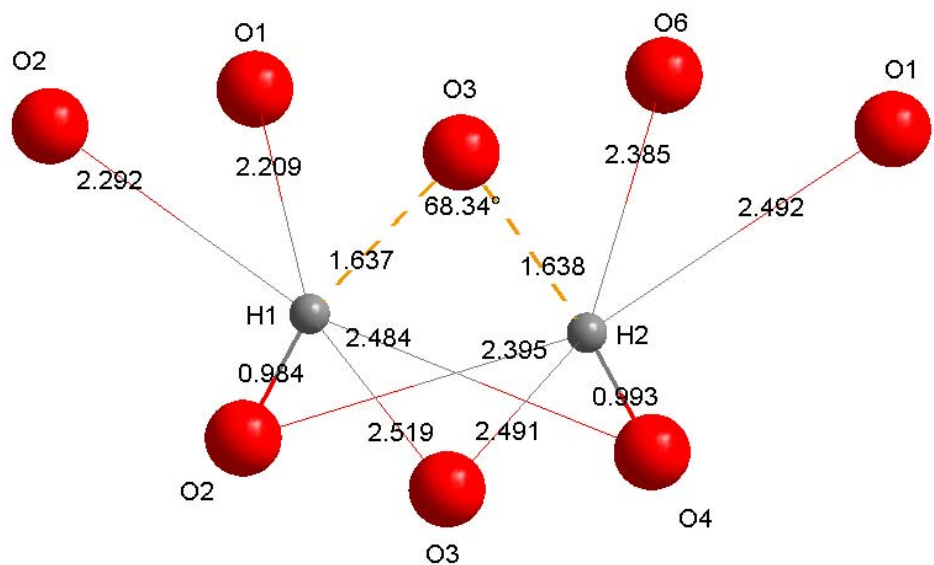

**(b)**

**Figure 7**

Table 1. Results of structural optimization at 0.1 MPa

|  | Calculation | Experiment [†] | Experiment [‡] | Experiment [*] |
|---|---|---|---|---|
| $a$(Å) | 7.9227 | 7.8620(3) | 7.8604(7) | 7.8678(4) |
| $c$(Å) | 9.6363 | 9.5752(3) | 9.5702(8) | 9.5771(5) |
| $V_o$(Å$^3$) | 523.8 | 512.56(3) | 512.08 | 513.43(4) |
| $B_o$(GPa) | 96.2 | 97.5(4) | 105 | 102.9(28) |
| $B'$ | 4.73 | 5.97(14) | 4 | 6.4(3) |

[†] Crichton and Ross[11]

[‡] Kudoh et al.[10] for the chemical composition $Mg_{6.99} Si_{1.99} H_{6.06}O_{14}$, which is slightly different from the ideal one. The authors ascribe this difference to the different total weight of $H_2O$.

[*] Holl et al[12] for iron bearing Phase A ($Mg_{6.85} Fe_{0.14} Si_{2.00} H_6O_{14}$).

Table. 2 The comparison between the observed and computed fractional coordinates of atoms for Phase A at the ambient experimental volume (512.62Å³, the calculated pressure at this volume is 2.2 GPa) and at a volume 474.2Å³ for experimental pressure of 9.4 GPa.

|  | Ambient volume 512.62 Å³ ($P_{calc}$ = 2.2 GPa ) | | $P_{exp}$ = 9.4 GPa (V = 474.2Å³) | |
|---|---|---|---|---|
|  | Experimental | Calculated | Experimental | Calculated |
| Si1 z | 0.1741 | 0.1745 | 0.1829 | 0.1758 |
| Si2 z | 0.4018 | 0.4015 | 0.4092 | 0.4017 |
| Mg1 x | 0.3722 | 0.3714 | 0.3757 | 0.3671 |
| y | 0.4547 | 0.4553 | 0.4602 | 0.4523 |
| z | 0.3857 | 0.3855 | 0.3901 | 0.3832 |
| Mg2 x | 0.2252 | 0.2236 | 0.2206 | 0.213 |
| y | 0.2438 | 0.2443 | 0.2450 | 0.2383 |
| z | 0.1127 | 0.1124 | 0.1224 | 0.1137 |
| Mg3 z | 0.1029 | 0.1016 | 0.1130 | 0.103 |
| O1 x | 0.2001 | 0.2019 | 0.2097 | 0.1993 |
| Y | 0.0274 | 0.0278 | 0.0330 | 0.022 |
| Z | -0.0240 | -0.0229 | -0.0229 | -0.0213 |
| O2 x | 0.4766 | 0.4755 | 0.4700 | 0.4686 |
| Y | 0.0988 | 0.0980 | 0.0985 | 0.0916 |
| Z | 0.4811 | 0.4835 | 0.4815 | 0.4847 |
| O3 x | 0.4538 | 0.4492 | 0.4486 | 0.4365 |
| Y | 0.2947 | 0.2866 | 0.2873 | 0.272 |
| Z | 0.2320 | 0.2339 | 0.2517 | 0.2376 |
| O4 x | 0.1704 | 0.1682 | 0.1604 | 0.1518 |
| Y | 0.4367 | 0.4371 | 0.4370 | 0.4302 |
| Z | 0.2398 | 0.2395 | 0.2549 | 0.2416 |
| O5 z | 0.0000 | 0.0000 | 0.0000 | 0.0000 |
| O6 z | 0.2323 | 0.2311 | 0.2401 | 0.2296 |
| H1 x | 0.4550 | 0.4555 |  | 0.4529 |
| Y | 0.1610 | 0.1665 |  | 0.1691 |
| Z | 0.4050 | 0.4042 |  | 0.4116 |
| H2 x | 0.0380 | 0.0255 |  | 0.0121 |
| Y | 0.3610 | 0.3478 |  | 0.3406 |
| z | 0.2400 | 0.2376 |  | 0.2444 |